\documentclass{article}

\PassOptionsToPackage{numbers, compress}{natbib}
\usepackage[final]{neurips_2019_ml4ps}
\usepackage[utf8]{inputenc} % allow utf-8 input
\usepackage[T1]{fontenc}    % use 8-bit T1 fonts
\usepackage{hyperref}       % hyperlinks
\usepackage{url}            % simple URL typesetting
\usepackage{booktabs}       % professional-quality tables
\usepackage{amsfonts}       % blackboard math symbols
\usepackage{nicefrac}       % compact symbols for 1/2, etc.
\usepackage{microtype}      % microtypography
\usepackage{amsmath}
\usepackage{graphicx} %package to manage images
\graphicspath{ {./Fig/} }
\usepackage{subfig}
\title{Accelerating Least Squares Imaging Using Deep Learning Techniques}
\author{%
  Janaki Vamaraju\\
  John A. and Katherine G. Jackson School of Geosciences, \\
  The University of Texas at Austin \\
   \AND
   Jeremy Vila \\
   Shell International Exploration and Production Inc. \\
   \AND
   Mauricio Araya-Polo \thanks{Mauricio Araya-Polo is now at Total. All contributions to this work were made while at Shell Global Solutions US Inc.} \\
   Shell Global Solutions US Inc. \\
   \AND
   Debanjan Datta \\
   Shell International Exploration and Production Inc. \\
   \AND
   Mrinal K. Sen \\
   John A. and Katherine G. Jackson School of Geosciences, \\
   The University of Texas at Austin \\
}
\begin{document}

\maketitle

\begin{abstract}
    Wave equation techniques have been an integral part of geophysical imaging workflows to investigate the Earth's subsurface. Least-squares reverse time migration (LSRTM) is a linearized inversion problem that iteratively minimizes a misfit functional as a function of the model perturbation. The success of the inversion largely depends on our ability to handle large systems of equations given the massive computation costs. The size of the system almost exponentially increases with the demand for higher resolution images in complicated subsurface media. We propose an unsupervised deep learning approach that leverages the existing physics-based models and machine learning optimizers to achieve more accurate and cheaper solutions. We compare different optimizers and demonstrate their efficacy in mitigating imaging artifacts. Further, minimizing the Huber loss with mini-batch gradients and Adam optimizer is not only less memory-intensive but is also more robust. Our empirical results on synthetic, densely sampled datasets suggest faster convergence to an accurate LSRTM result than a traditional approach.
\end{abstract}

\section{Introduction}

\begin{figure}%
    \centering
    \subfloat{{\includegraphics[width=6cm]{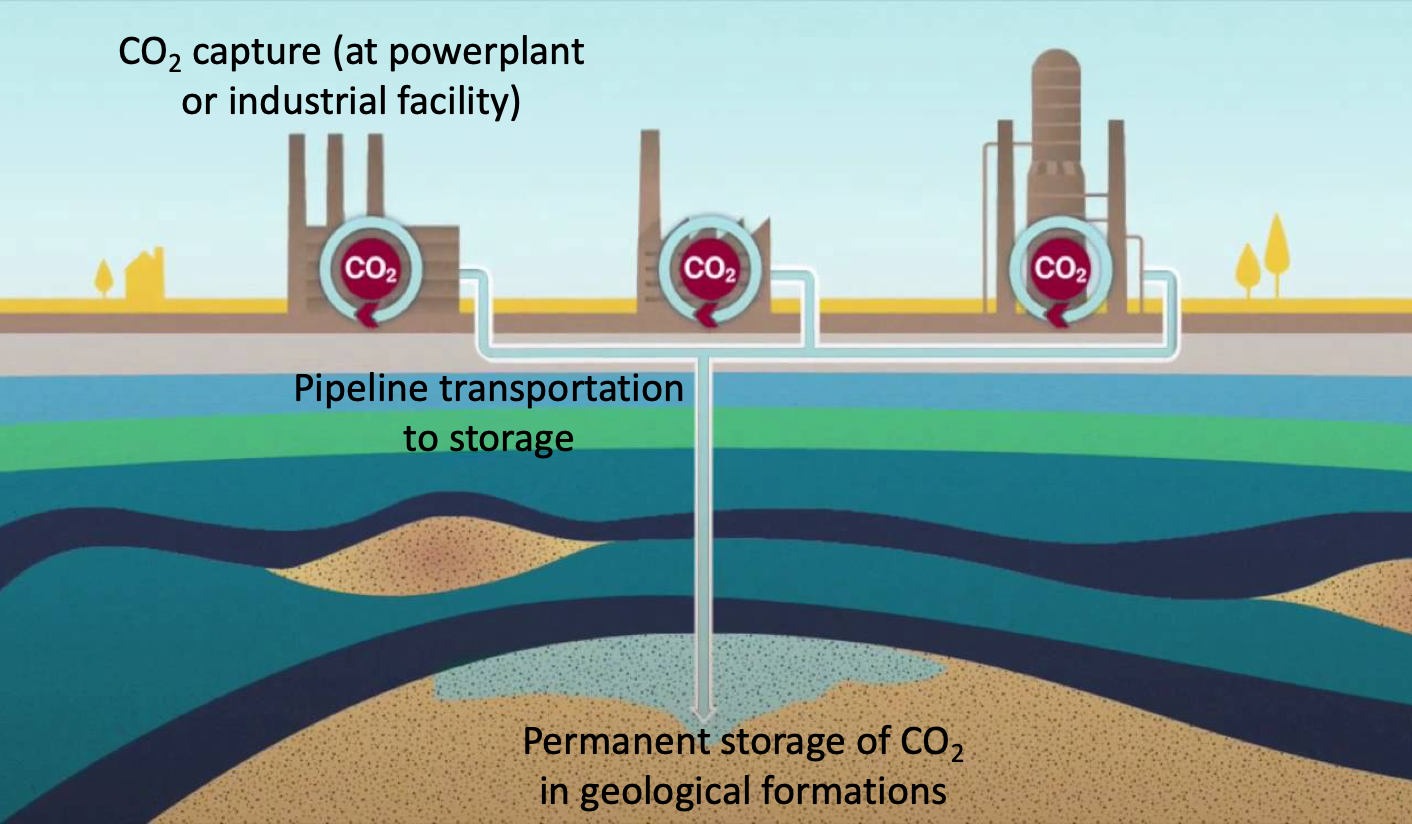} }}
    \caption{Geophysical imaging of geological formations for enhanced carbon capture and storage}%
    \label{fig:Fig0}%
\end{figure}

 Imaging techniques are used in geophysics to produce images of the Earth's subsurface at diverse length scales. These techniques are recently being adopted for monitoring subsurface geological formations that are used for carbon capture and storage (Figure \ref{fig:Fig0}). $CO_{2}$ sequestration is one of the  solutions to the increasing greenhouse gas emissions that cause global warming and climate change. In this context, wave equation based migration algorithms \cite[]{baysal1983reverse,mcmechan1983migration} have been an integral part of imaging workflows to invert for subsurface properties at greater depths. Among them, least-squares reverse time migration (LSRTM) \cite[]{nemeth1999least,kuhl2003least} is the most popular migration method due to its ability to image complex subsurface areas with large computational resources such as graphic processing units (GPUs). LSRTM can be seen as a linear inverse problem based on the acoustic wave equation. The goal is to invert for an earth model that represents rock properties to fit the recorded surface data. The imaging condition or the gradient calculation in LSRTM is to take the zero-lag of the cross-correlation between the reverse-time-extrapolated receiver wavefield and the forward-time-extrapolated source wavefield \cite[]{claerbout1971toward}. When the data is subjected to aliasing, truncation or noise, the adjoint operator can degrade the resolution of the final migrated image. LSRTM obtains an approximate image of the model perturbation by iteratively minimizing the cost function. The LSRTM, however, is an expensive replacement and its ability to migrate surface data depends on the accuracy of the velocity model and adequate preconditioning \cite[]{rickett2003illumination,guitton2004amplitude}, and needs additional regularization terms for successful damping of artifacts \cite[]{xue2015seismic,wu2016least}. At each iteration of LSRTM, Born forward modeling and adjoint operators are applied, which makes the computational cost extremely high. 

Under the paradigm of theory guided data science, many researchers in the geophysics community are looking at ways of combining physics and machine learning \cite[]{calderon1997hopfield,vamaraju2019unsupervised,araya2017automated,hansen2017efficient,araya2018deep, adler2019deep,biswas2019pre,richardson2018seismic}. These works, however were not extended to least-squares migrations. In this paper, we implement LSRTM using a deep learning approach and adopt strategies from data science to reduce computational costs and accelerate convergence.The feasibility of LSRTM is examined in combination with mini-batch gradients and deep learning optimizers such as the Hopfield neural networks (HNN), adaptive moment estimation (Adam) and Limited memory BFGS (L-BFGS). Mini-batch gradients help in reduction of cross-talk \cite[]{friedlander2012hybrid,van2013fast,yang2018mini} and the deep learning optimizers can help mitigate acquisition footprints that are caused by the lack of shot data. This not only achieves faster convergence through iterations but also generates geologically consistent models. The computation cost is further reduced by using a subset of total shot data for each iteration. Implementing LSRTM in a deep learning framework (Pytorch or Tensorflow) enables us to experiment with machine learning loss functions and regularizations. The automatic differentiation capability of the software can be used to calculate the gradient of the cost function. We further minimize the Huber loss function to improve the efficiency of LSRTM. We apply the techniques to a 2D synthetic model and show improvement over conventional LSRTM baselines. The proposed methodology achieves higher spatial resolution according to quantitative evaluation metrics. 

\section{Methodology}

The goal of least-squares migration is to invert for the earth's  surface  reflectivity model (model perturbation indicating rock properties) \textbf{m} to fit the recorded data, $\mathbf{d_{0}}$:

\begin{align*}
    C(\mathbf{m}) = \frac{1}{2}\left \| \mathbf{d_{0}}-\mathbf{G}\mathbf{m} \right \|^{2},    (1)
\end{align*}
where $C$ is the cost/loss function to be minimized and \textbf{G} is the linearized Born modeling (scattering) operator that requires a background velocity model. This background velocity model is known a-priori from other velocity analysis methods. If $\mathbf{G^{T}}\mathbf{G}$ is invertible, the least-squares solution for equation $(1)$ can be written as:

\begin{align*}
    \mathbf{m} =(\mathbf{G^{T}}\mathbf{G})^{-1}\mathbf{G^{T}}\mathbf{d_{0}},  (2)
\end{align*}
where $\mathbf{G^{T}}$ is the migration operator and $\mathbf{G^{T}}\mathbf{G}$ is the Hessian matrix $\mathbf{H}$. The key to LSRTM is to obtain the inverse of $\mathbf{H}$; however the computational cost and storage of $\mathbf{H}$ are not feasible for realistic problems. Alternatively, different approximations, such as gradient based iterative approaches \cite[]{schuster1993least,nemeth1999least,tang2009target} are pursued. In this paper, we use PyTorch \cite[]{paszke2017automatic} for the above implementation. The Born modeling based wave equations can be implemented as a recurrent neural network (RNN) which is similar to the workflow proposed by \cite{richardson2018seismic}. The same operations are applied in each cell of an RNN, but the data
that the operations act upon changes. Each cell applies the finite-difference convolution operation to propagate forward one time step, by taking the state from the previous cell (the displacement wavefield at adjacent time steps and auxiliary wavefield) and the source amplitude as inputs, and producing the updated state vectors and the current wavefield as outputs. Although, automatic differentiation can be used to calculate the gradient of the cost function, we use the traditional adjoint state method to save GPU memory costs. Easy chaining of operations in PyTorch, enables users to create customized loss functions or apply data normalizing operations. We generate observed data with the full finite difference wave propagation operator. LSRTM intrinsically is an under-determined inverse problem. There are always a limited number of receivers that can cover the subsurface and the source function is a band limiting signal. This can lead to incomplete data in terms of spatial coverage and frequency content. Here, we use the Huber loss function \cite[]{huber1992robust,guitton2003robust} to improve the resolution and robustness of LSRTM:

\begin{align*}
     C(\mathbf{m}) =  \frac{1}{N} \sum_{i}^{N}C_{i}(\mathbf{m}) \: \text{where} \: C_{i}(\mathbf{m}) = \Bigg\{ \begin{matrix}
\frac{1}{2}\left \| \mathbf{d_{0}}-\textbf{G}\textbf{m} \right \|^{2} \: \: \: \: \: \: \: \:   \text{if} \: \left |\mathbf{d_{0}}-\textbf{G}\textbf{m}  \right | \leq \varepsilon  \\ \varepsilon \left |\mathbf{d_{0}}-\textbf{G}\textbf{m} \right |-\varepsilon /2 \: \: \: \: \: \: \: \: \: \: \: \: \: \: \:  \text{otherwise}
\end{matrix},   (3)
\end{align*}
where $i$ is the corresponding shot record for $i$-th source and $N$ is the total number of shots.  The parameter $\varepsilon$ defines a threshold based on the distance between target and prediction. For the example in this paper, we use a default value of 1 for $\varepsilon$. Conventionally, the LSRTM cost function is evaluated at each iteration using all the shot data that is available. The gradient calculation can be very expensive when the number of shots are large (especially in 3D surveys). To reduce the computation costs and to reap the benefits of stochastic optimization, mini-batch gradient methods take a subset of entire shots to construct the objective function and update the model:

\begin{align*}
    C(\mathbf{m})\approx C_{k}^{B}(\mathbf{m}) = \frac{1}{\left |B  \right |} \sum_{i\in B}C_{i}(\mathbf{m}),    (4)
\end{align*}
where $B$ is  a  subset  of  total  shots,  and $\left |B  \right |$ is  its  size. 
This performs frequent updates with reasonable variance and is faster to converge. The fluctuation of the objective function is not severe. We first divide all the shot data into sequential mini-batches and then randomly shuffle them. Although mini-batches offer a more scalable solution, they can introduce migration artifacts if the data is not sampled effectively. Therefore, we need more efficient optimization algorithms to produce migrated images with higher resolution. First-order gradient based optimization algorithms can be very slow to converge to high accuracy solutions, specially in noisy large-scale datasets. We combine mini-batch gradients with various optimizers such as the Hopfield neural networks (HNN), adaptive moment estimation (Adam) and Limited memory BFGS (L-BFGS). These techniques not only reduces the number of forward solves but also enhances the accuracy of inverted results when compared to traditional LSRTM algorithms (using gradient descent (GD)). We provide qualitative and quantitative comparisons of the above presented methods against conventional LSRTM baseline migrated images. Note that, no preconditioning was used for either conventional LSRTM or for the proposed mini-batch approach.

\section{Results and Discussion}
\begin{figure}%
    \centering
    \subfloat{{\includegraphics[width=\columnwidth]{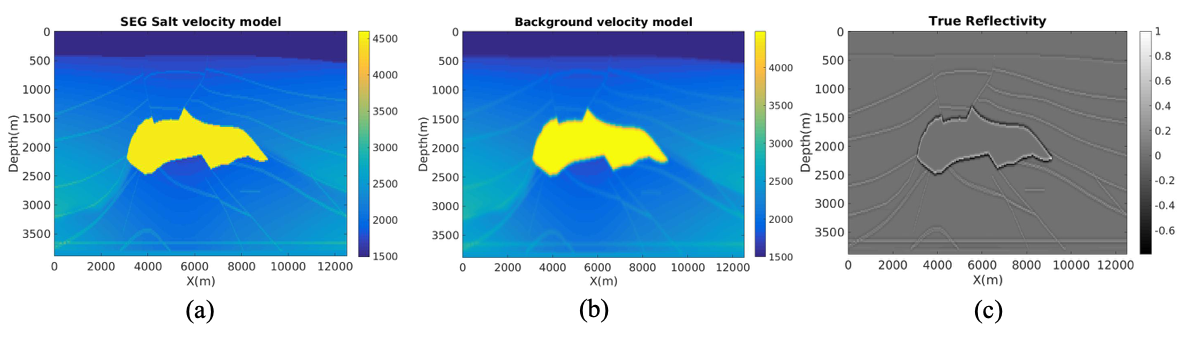} }}
    \caption{SEG/EAGE salt model (a) True velocity (b) Smooth migration velocity model (c) True reflectivity}%
    \label{fig:Fig1}%
\end{figure}

\begin{figure}%
    \centering
    {{\includegraphics[width=\columnwidth]{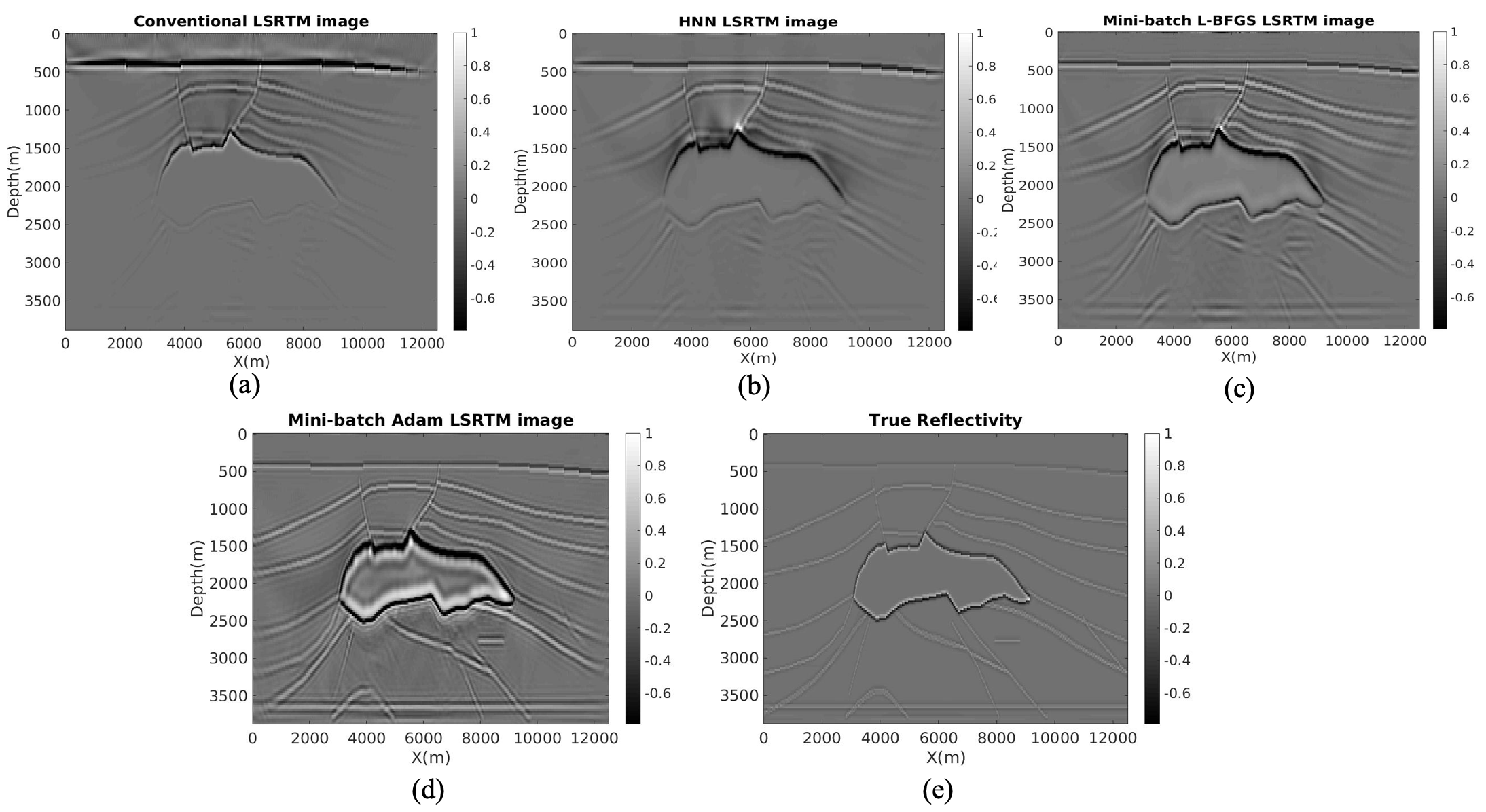} }}
    \caption{SEG/EAGE salt model migrated images (a) Conventional LSRTM image (MSSIM=0.7755) (b) HNN LSRTM image (MSSIM=0.79) (c) Mini-batch L-BFGS LSRTM image (MSSIM=0.8026) (d) Mini-batch Adam LSRTM image (MSSIM=0.8255) (e) True reflectivity}%
    \label{fig:Fig2}%
\end{figure}

\begin{figure}%
    \centering
    {{\includegraphics[width=\columnwidth]{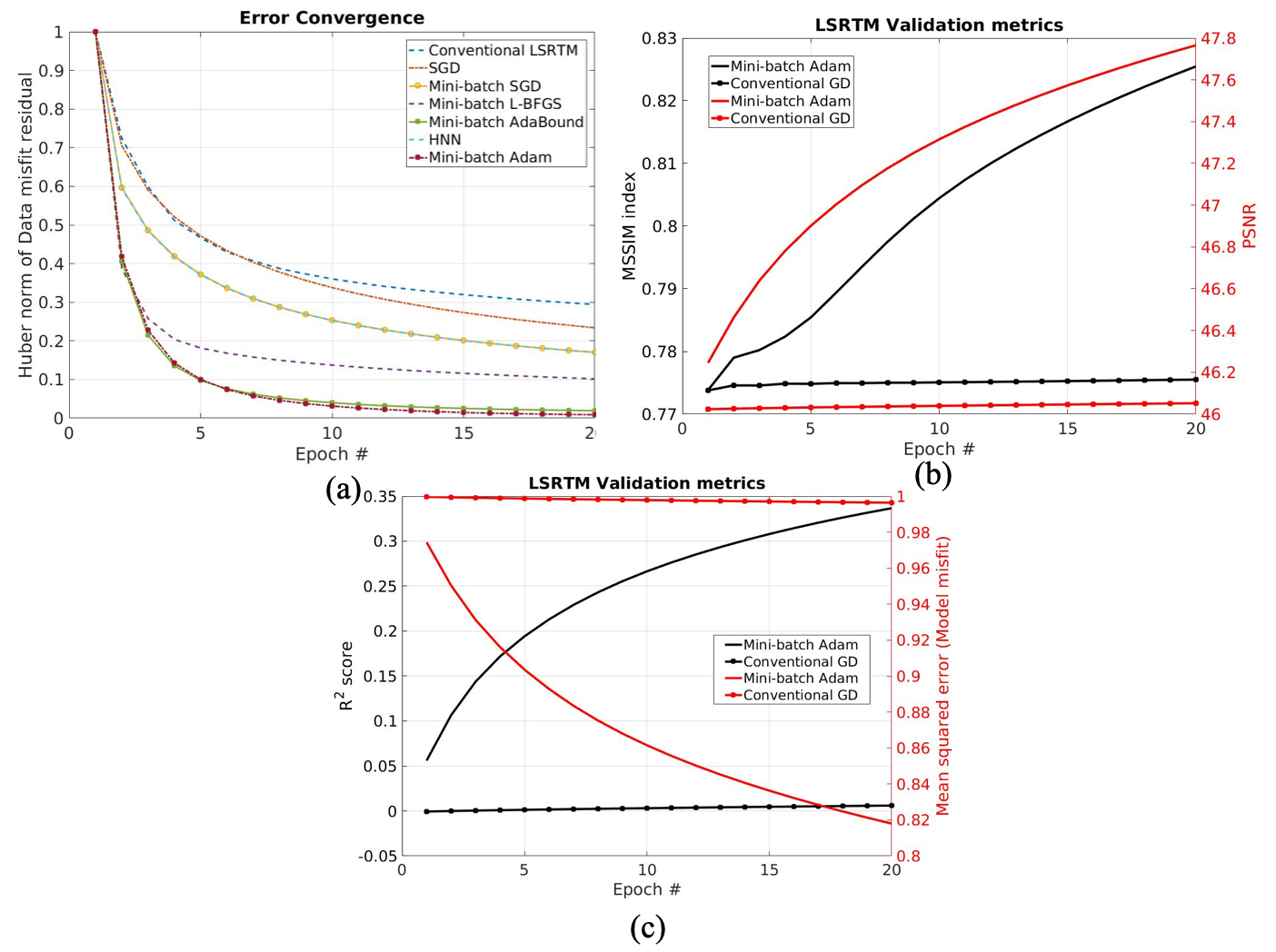} }}
    \caption{Comparison of (a) Error convergence (b) MSSIM and PSNR (c) $R^{2}$  and MSE metrics for conventional LSRTM and mini-batch Adam LSRTM images}%
    \label{fig:Fig4}%
\end{figure}

The synthetic example is based on a $2D$ slice from the SEG/EAGE $3D$ salt model \cite[]{aminzadeh1996three}. This model is 4 km in depth and 12.5 km wide. We set 160 shots and 160 receivers per shot both deployed on the surface. The peak frequency of the source Ricker wavelet is 6 Hz. Figures \ref{fig:Fig1}(a), \ref{fig:Fig1}(b) and \ref{fig:Fig1}(c) show the true velocity model, the background smooth velocity model and the true reflectivity respectively. Figure \ref{fig:Fig2}(a) shows a conventional LSRTM image using GD after 20 epochs. The reflectivity is still far from the true reflectivity and the bottom of the salt body is not very clear. Figure \ref{fig:Fig2}(b) shows the migrated image from HNN which is slightly improved with a mean structural similarity index (MSSIM) of 0.79. Figures \ref{fig:Fig2}(c) and \ref{fig:Fig2}(d) show the migrated images from L-BFGS and Adam optimizers respectively. When combined with mini-batch training, not only the true amplitudes are better recovered but also the salt bottom is now clearly resolved. The reflectors at the bottom of the image are enhanced and there is no loss of continuity. The number of shots in each mini-batch and the step-size for model updates are the two important hyperparameters. We randomly reserve twenty-five percent of all the shots that evenly cover the entire survey. This development dataset is used to evaluate the cost function for 25 combinations of batch-sizes and learning rates (Grid-search optimization). The  batch  size  and learning  rate  that  correspond to  lowest  cost function  value  is chosen for each optimizer. For 40 shots and Adam optimizer, the best learning rate and batch-size is $1e^{-7}$ and $4$ respectively. The migrated images have an MSSIM index greater than 0.8 and the migration artifacts are removed. Figure \ref{fig:Fig4}(a) shows the error convergence or the cost function calculated for each optimizer. Adam and AdaBound converge within 6 epochs and the cost function becomes flat beyond 10 epochs. However, the image from Adam (MSSIM=0.8255) has a higher MSSIM than AdaBound (MSSIM = 0.8157). The performance of L-BFGS is comparable but converges in slightly more number of epochs than Adam. Figures \ref{fig:Fig4}(b) and \ref{fig:Fig4}(c) plot the other validation metrics for a quantitative evaluation. Using the Adam optimizer, at the end of 20 epochs, MSSIM index reaches a value of 0.8255 from 0.77. Compared to Adam, conventional LSRTM shows almost no improvement from the initial image. Additional preconditioning and filtering is usually needed to greatly improve the image. The final $R^{2}$ score of the Adam migrated image is raised to 0.34 from an initial 0.06. The mean squared error (MSE) value is reduced to 0.825 and the peak signal-to-noise ratio (PSNR) is improved to 47.78 over 20 epochs. Note that the PSNR value is measured in log scale. The performance trend is overall positive. 
\section{Conclusions}
We present a deep learning approach for LSRTM by adopting strategies from data-science to accelerate convergence. In a time-domain formulation, mini-batch gradients can reduce the computation cost by using a subset of total shots for each epoch. The Adam optimizer, combined with the Huber loss and mini-batch gradients resulted in significantly faster convergence than when using conventional gradient descent optimizer with the cost and gradient calculated using the entire dataset. We apply the techniques to the SEG/EAGE 3D salt model and show improvements over conventional LSRTM baseline. The described methods mitigate artifacts that arise from limited aperture, low subsurface illumination, and cross-correlation noise. Effects that irregular sampling has on the proposed mini-batch approach remains future work.

\newpage
\bibliographystyle{unsrt}
\bibliography{refs}

\end{document}